\documentclass[twocolumn,showpacs,preprintnumbers,amsmath,amssymb]{revtex4}
\usepackage{epsf}
\begin{document}
 
\title{Resonant activation in a nonadiabatically driven optical lattice}
  
\author{R. Gommers, P. Douglas, S. Bergamini, M. Goonasekera, 
P.H. Jones and F. Renzoni}
  
\affiliation{Departement of Physics and Astronomy, University College London,
Gower Street, London WC1E 6BT, United Kingdom}
  
\date{\today}
  
\begin{abstract}
We demonstrate the phenomenon of resonant activation in a non-adiabatically 
driven dissipative optical lattice with broken time-symmetry. The resonant 
activation results in a resonance as function of the driving frequency
in the current of atoms through the periodic potential. We demonstrate that 
the resonance is produced by the interplay between deterministic driving and 
fluctuations, and we also show that by changing the frequency of the driving
it is possible to control the direction of the diffusion.
\end{abstract}
                                                                                
\maketitle

The problem of the escape of a Brownian particle out of a potential well, 
first characterized by Kramers \cite{kramers}, plays a central role in the
description of many processes in physics, chemistry and biology 
\cite{rmp,melnikov}. Kramers' law predicts an escape rate of the form
$\exp{(-U/kT)}$, where $U$ is the depth of the well, $T$ the temperature and
$k$ the Boltzmann constant. In the case of a nonadiabatically driven Brownian
particle the aforementioned scenario may change significantly, and much work 
has been devoted to the study of this non-equilibrium phenomenon
\cite{devoret,mannella,dykman,luchinsky1,luchinsky2,lehmann,chaos,soskin}. It 
has been shown that in the presence of nonadiabatic driving the lifetime of the
particle in the potential well can be significantly reduced, a phenomenon 
named resonant activation.

Resonant activation was firstly observed in a current-biased Josephson 
tunnel junction \cite{devoret}, and more recently for a Brownian particle 
optically trapped in a double well potential \cite{chaos}. Resonant 
activation has also been theoretically studied for Brownian particles in
periodic potentials, a configuration relevant for the realization of Brownian
motors \cite{ratchet}. Also in this case the nonadiabatic driving may result 
in a significant enhancement of the activation rate. Furthermore it has been
predicted that, whenever the spatio-temporal symmetry of the system is broken,
the resonant activation gives rise to {\it resonant} rectification of 
fluctuations \cite{dykman,goychuk,luchinsky2}. Indeed, very recently 
Brownian motors have been realized with nonadiabatically driven Brownian
particles \cite{weiss,renzoni1}. However in those works the resonant nature
of the rectification mechanism, and the underlying resonant activation process,
have not been investigated. In particular, the possibility to control the 
amplitude and direction of the diffusion by changing the frequency of the 
driving field, as predicted by theoretical work \cite{dykman,soskin,luchinsky2},
has not yet been demonstrated.

In this Letter we demonstrate experimentally the phenomenon of resonant 
activation in a Brownian motor by using cold atoms in a driven dissipative 
optical lattice as a model system. Indeed, we observe the appearance of a 
resonance while monitoring the current of atoms through the lattice as a 
function of the driving frequency. We show that this resonance is due to the 
interplay between fluctuations and deterministic driving, and we demonstrate 
that by varying the driving frequency it is possible to reverse the current
direction.

In our experiment we use cold atoms in a dissipative optical lattice
\cite{robi}, in which the atom-light interaction determines both the periodic
potential for the atoms and the dissipation mechanism which leads to a friction
force -- the so called Sisyphus cooling -- and to fluctuations in the atomic 
dynamics. This system offers the significant advantage of being easily tunable 
over a wide range of parameters: the potential depth, the fluctuations level 
and the parameters (frequency and amplitude) of the driving force can be 
varied and carefully controlled over a large interval of values. This is an
essential feature for the investigation presented here.

Before presenting the experimental results, we analyze our system
through numerical simulations. For the sake of clarity, our theoretical 
analysis will be limited to the simplest one-dimensional configuration in 
which Sisyphus cooling takes place: a $J_g=1/2\to J_e=3/2$ atomic transition
and two counterpropagating laser fields with orthogonal linear polarizations 
-- the so called lin$\perp$lin configuration. The light interference pattern 
results in two optical potentials $U_{\pm}$ for the atoms, one for each 
ground state $|\pm\rangle$, in phase opposition: 
$U_{\pm}=U_0\left[-2\pm \cos2kz\right]/2$, where $z$ is the atomic 
position along the axis $Oz$ of light propagation, $k$ the laser field 
wavevector and $U_0$ the depth of the optical potential.  The stochastic
process of optical pumping transfers the atom from one ground state to the 
other one, changing in this way the optical potential experienced by the atom. 
This stochastic process results in a friction force and produces fluctuations 
in the atomic dynamics \cite{robi}.  The departure rates
$\gamma_{\pm\to \mp}(z)$ from the $|\pm\rangle$
states can be written in terms of the photon scattering rate 
$\Gamma'$ as $\gamma_{\pm\to \mp}= \Gamma' (1\pm \cos 2kz) /9$ \cite{robi}. 
It appears therefore that the amplitude of the fluctuations can be 
quantitatively characterized by the photon scattering rate $\Gamma'$, which is 
an experimentally accessible parameter.

To study the phenomenon of resonant activation we need to drive the atoms
with a zero-mean oscillating force. We consider here an ac force consisting 
of two harmonics, $A_1\cos{(\omega t)}$ and $A_2\cos{(2\omega t-\pi /2 )}$,
so that the resonant activation should lead, following the breaking of the 
time-symmetry of the system, to a resonant generation of a current, as 
predicted by theoretical work \cite{dykman,luchinsky2,general}. 
In the numerical work it is obviously possible to "apply" directly an 
homogeneous ac force to the atoms, by simply including the appropriate terms
in the equation of motion. On the contrary, in the experiment this is not
possible, and forces can be applied only by phase modulating the lattice
beams. For consistency, we follow the same approach in the theoretical 
analysis, and we consider a phase-modulation $\alpha(t)$ of one of the lattice
beams of the form
\begin{equation}
\alpha(t) = \alpha_0 \left[ A_1 \cos{(\omega t})+
\frac{1}{4}A_2\cos{(2\omega t-\pi/2})\right] ~.
\label{alpha}
\end{equation}
In this way in the accelerated frame in which the optical potential is 
stationary the atoms experience an inertial force 
\begin{equation}
F(t) = \frac{m\omega^2\alpha_0}{2k} \left[ A_1\cos{(\omega t})+
A_2\cos{(2\omega t-\pi/2})\right] ~,
\label{force}
\end{equation}
where $m$ is the atomic mass and $k$ the laser field wavevector.

To study the atomic dynamics in the presence of the nonadiabatic driving, 
we follow the same procedure developed to investigate laser cooling processes
in (undriven) optical lattices. The Fokker-Plank-type equation for the undriven
system, and the Monte Carlo technique to derive the atomic trajectories have
been discussed in detail in Refs. \cite{robi,petsas}. The generalization of
that method for the driven system of interest here is straighforward, and
consists in the inclusion of a time-dependent shift $\alpha (t)$ (see Eq. 
(\ref{alpha})) in the relative phase between the two laser fields generating 
the optical lattice.

\begin{figure}[ht]
\begin{center}
\mbox{\epsfxsize 3.in \epsfbox{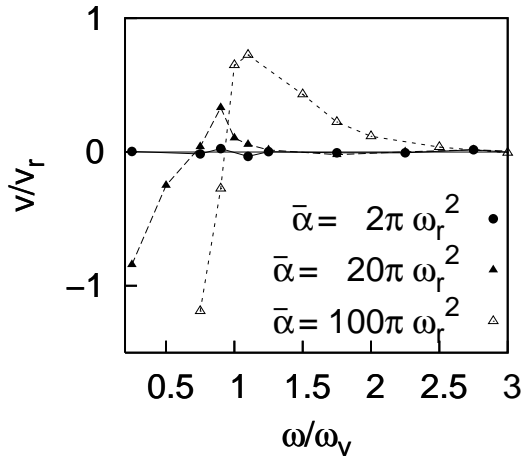}}
\end{center}
\caption{Results of Monte Carlo simulations for a sample of $n=10^4$ atoms 
in a 1D lin$\perp$lin optical lattice. The mean atomic velocity $v$,
rescaled by the atomic recoil velocity $v_r=\hbar k/m$, is shown
as a function of the driving frequency $\omega$, for different amplitudes of
the ac force. The driving frequency $\omega$ is expressed in terms of 
the vibrational frequency $\omega_v$ of the atoms at the bottom of the 
potential wells. Parameters of the calculation are: $\Gamma'=10\omega_r$,
$U_0=100\cdot E_r$, with $E_r$ and $\omega_r$ the recoil energy and frequency,
respectively. The amplitude of the two harmonics of the force are: 
$A_1=1$, $A_2=4$.
}
\label{fig1}
\end{figure}

Through Monte Carlo simulations, we calculated  the mean atomic velocity $v$ 
as a function of the frequency $\omega$ of the driving, for different 
amplitudes of the ac force. The results of our calculations are shown in 
Fig.~\ref{fig1}. For each data set, in order to keep constant the amplitude 
$F_0=m\omega^2\alpha_0/2k$ of the ac force
(see Eq.~(\ref{force})) we varied the amplitude $\alpha_0$ of 
the phase modulation according to $\alpha_0=\bar{\alpha}/\omega^2$, with  
$\bar{\alpha}$ constant for a given $F_0$. This is the same procedure used in
the experiment.

Figure \ref{fig1} shows clearly the appearance of a resonance in the current 
amplitude as a function of the driving frequency. 
%
%
Indeed, for weak driving the current is negligible. At larger amplitude of
the driving the current differs significantly from zero, and shows a well 
defined resonance. The resonance is observed in the regime of non-adiabatic
driving, i.e. for driving frequencies of the order of or exceeding the
vibrational frequency. The numerical simulations also show that by
changing the driving frequency it is possible to reverse the current
direction, as predicted by the general theory 
\cite{dykman,soskin,luchinsky2}. We note that we carried out numerical simulations
for two different ratios of the force harmonics' amplitude: $A_1/A_2=1/4$
(Fig.~\ref{fig1}) and $A_1/A_2=1$ (not shown). The two sets of calculations
evidenced the same qualitative behaviour.

\begin{figure}[ht]
\begin{center}
\mbox{\epsfxsize 3.in \epsfbox{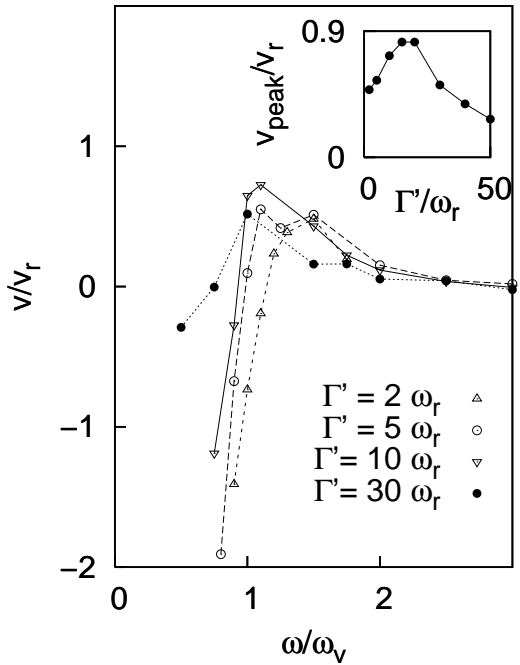}}
\end{center}
\caption{Results of Monte Carlo simulations for a sample of $n=10^4$ atoms
in a 1D lin$\perp$lin optical lattice. The mean atomic velocity $v$,
is shown as a function of the driving frequency $\omega$, for different 
values of the scattering rate $\Gamma'$. In the inset, the mean atomic 
velocity $v_{\rm peak}$ at the maximum of the resonance is reported as
a function of the scattering rate.  The depth of the optical 
potential is $U_0=100\cdot E_r$; the ac force amplitude corresponds to
$\bar{\alpha} = 100\pi\omega_r^2$. The amplitudes of the two harmonics of the 
force are : $A_1=1$, $A_2=4$.  The lines are guides for the eye.}
\label{fig2}
\end{figure}

To determine the nature of the resonance we focus our attention on the
range of frequencies of the ac fields corresponding to non-adiabatic driving.
This regime is illustrated in Fig.~\ref{fig2}, where the resonance in the
atomic current as a function of the driving frequency is shown for a given 
amplitude of the ac field and at different values of the scattering 
rate.  From Fig.~\ref{fig2}
it appears that the amplitude of the resonance shows a non-monotonic dependence
on the scattering rate $\Gamma'$: at small $\Gamma'$ the resonance 
amplitude increases with $\Gamma'$, then reaches a maximum and finally at
larger values of $\Gamma'$ decreases. The constructive role played by the noise
at low levels of $\Gamma'$ shows that the resonance observed
in the numerical simulations is determined by the interplay between the 
applied ac forces and the random fluctuations, which results in the 
rectification of the latter ones. This is at variance with the behavior 
observed at frequencies somewhat below $\omega_v$ where the magnitude
of the (negative) current is decreased by an increase of
the scattering rate, a behaviour characteristic of deterministic 
rectification of forces. The conclusion of our numerical analysis is therefore
that the resonant activation phenomenon should be observable in a dissipative
optical lattice, and should result in a resonance in the atomic current
as a function of the driving frequency.

The experiment is a direct implementation of what described in the theoretical
analysis. Instead of using a 1D optical lattice, as in the numerical work,
we use a 3D lattice, which offers the significant advantage to confine the
atoms in the three directions. This reduces the losses of atoms
from the lattice during the experiment.

The experimental set-up is the same as the one used in Ref.~\cite{renzoni1},
and consists of four linearly-polarized laser beams arranged in the so-called
umbrellalike configuration \cite{petsas}. One laser beam propagates in the 
$z$-direction. This is the beam which will be phase-modulated. 
The three other laser beams propagate in the opposite direction, 
and are arranged along the edges of a
triangular pyramid having the $z$-direction as axis, with the azimuthal angle
between each pair of beams equal to $2\pi/3$. 
For this lattice beam configuration, the interference of the laser fields
produces a periodic and spatially symmetric optical potential, with the
potential minima associated with pure circular ($\sigma^{+}$ or $\sigma^{-}$)
polarization of the light \cite{petsas}. For an atom with a $F_g=F\to
F_e=F+1$ transition, the optical potential consists precisely of $2F+1$
potentials, one for each ground state sublevel of the atom.

Cesium atoms are cooled and trapped in a magneto-optical trap (MOT). At a
given instant the MOT is turned off and the four lattice beams are turned on.
The atoms are left in this undriven optical lattice for 2 ms.
This is sufficient for the atoms to thermalize and reach equilibrium. Then
the phase modulation $\alpha (t)$ [see Eq.~(\ref{alpha})] 
is slowly turned on. The dynamics
of the atoms is studied with a charged-coupled device (CCD) camera. After
a short transient, the center-of-mass of the atomic cloud is observed to be
set into uniform motion along the $z$ axis, and a center-of-mass velocity $v$ 
is correspondingly derived. 

\begin{figure}[ht]
\begin{center}
\mbox{\epsfxsize 3.in \epsfbox{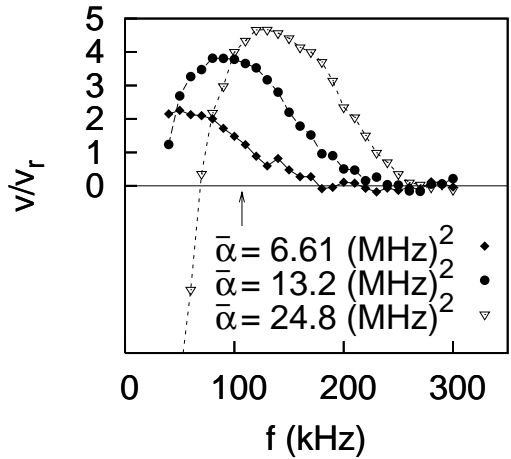}}
\end{center}
\caption{Experimental results for the average atomic velocity as a function
of the driving frequency, for different amplitudes of the driving force.
The optical potential is the same for all measurements and corresponds to
a vibrational frequency $\omega_v=2\pi\cdot 170$ kHz. 
The driving frequency satisfying the condition $2\omega=\omega_v$ is
indicated by an arrow.
The detuning $\Delta$ of the lattice from atomic resonance is 
$\Delta=11.1\Gamma$, where $\Gamma$ is the excited state 
linewidth.
The values for the
velocity are expressed in terms of the recoil velocity $v_r$, equal to
$3.52$ mm/s for the Cs D$_2$ line. The two harmonics of the force have equal
amplitude: $A_1=A_2=1$. The lines are guides for the eye.  }
\label{fig5}
\end{figure}

Results for the average atomic velocity
as a function of the driving frequency are shown in Fig.~\ref{fig5}
for different values of the amplitude of the driving. We clearly observe
the build-up of a resonance when the amplitude of the driving is progressively
increased. The resonance appears in the regime of non-adiabatic driving
($2\omega \gtrsim \omega_v$), and
a current reversal is observed on the low-frequency side of the resonance, in 
agreement with our simulations and with the general theory 
\cite{dykman,soskin,luchinsky2}. 

\begin{figure}[ht]
\begin{center}
\mbox{\epsfxsize 3.in \epsfbox{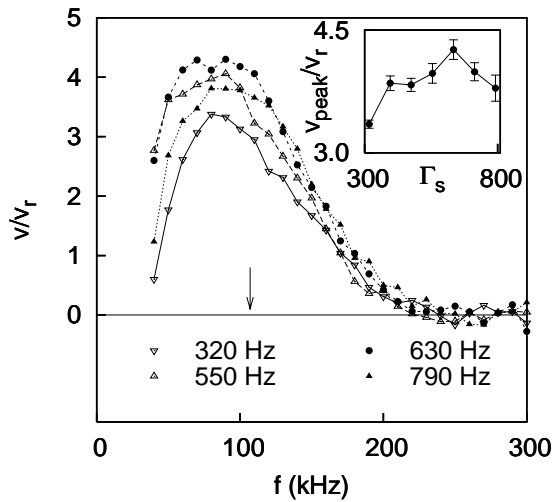}}
\end{center}
\caption{Experimental results for the average atomic velocity as a function
of the driving frequency, at different scattering rates. In the inset the 
resonance amplitude, i.e. the peak mean atomic velocity, is reported as a
function of the scattering rate.  The optical potential, and the driving 
force amplitude, are the same for all data sets, and they are characterized 
by $\omega_v=2\pi\cdot 170$ kHz  and $\bar{\alpha}= 13.3 ({\rm MHz})^2$.
The driving frequency satisfying the condition $2\omega=\omega_v$ is 
indicated by an arrow.
Different data sets correspond to different detunings $\Delta$, i.e., to 
different scattering rates as the optical potential is kept constant. The data
are labeled by the quantity $\Gamma_s=[\omega_v/(2\pi)]^2/\Delta$ which is
proportional to the optical scattering rate. The two harmonics of the force 
have equal amplitudes: $A_1=A_2=1$.  The lines are guides for the eye.}
\label{fig6}
\end{figure}

We examine now the dependence of the resonance on the scattering rate,
i.e. the fluctuations amplitude, with the aim to demonstrate that the observed 
resonance is indeed produced by the interplay of deterministic driving and
fluctuations. Figure \ref{fig6} shows our results for the average atomic
velocity as a function of the driving frequency, at different values of the
scattering rate. It clearly appears that the resonance amplitude shows a 
non-monotonic dependence on the scattering rate: the resonance initially
increases at increasing values of $\Gamma'$, then reaches a maximum and starts
decreasing at large values of the scattering rate.
This is in agreement with our numerical results and clearly demonstrates
that the resonance is determined by the interplay between deterministic 
driving and fluctuations and, due to the broken time-symmetry, results in the
rectification of the latter ones. 

In conclusion, in this work we demonstrated the phenomenon of resonant 
activation in a nonadiabatically driven dissipative optical lattice with
broken time-symmetry. Due to the broken symmetry 
of the system, the resonant activation results in a resonance in the current 
of atoms through the periodic potential. We demonstrated that the resonance 
is produced by the interplay between deterministic driving and fluctuations,
and we also showed that by changing the frequency of the driving it is 
possible to control the direction of the diffusion, as predicted by 
theoretical models. We notice that the rectification of fluctuations with 
non-adiabatically driven Brownian particles was already observed in previous 
work \cite{weiss,renzoni1}, but the resonant nature of the rectification
mechanism was not demonstrated. Our work therefore also establishes
experimentally the connection between resonant activation and resonant 
rectification of fluctuations, and confirms the theoretical predictions.
The present experimental realization, in which both deterministic and 
fluctuating forces originate from light fields, also shows the generality
of the phenomenon of resonant activation, which is not restricted to 
systems in which the fluctuations are of thermal origin, as usually 
considered in theoretical work.

We thank EPSRC, UK and the Royal Society for financial support.

\end{document}